\RequirePackage{fix-cm}
\documentclass[smallextended]{svjour3}       % onecolumn (second format)
\smartqed  % flush right qed marks, e.g. at end of proof
\usepackage{graphicx}
\usepackage{amsmath}
\usepackage{graphicx,psfrag,epsf}
\usepackage{enumerate}
\usepackage{apacite}
\usepackage{natbib}
\usepackage{url}
\usepackage{latexsym,amssymb}
\usepackage{epigraph,xcolor}

\usepackage{comment}
\excludecomment{pdfexclude}
\usepackage{booktabs}
\newcommand{\ra}[1]{\renewcommand{\arraystretch}{#1}}

\usepackage{multirow}
\usepackage{comment}
\usepackage{apalike}
\excludecomment{mysection}

\usepackage{setspace}
\begin{document}

\title{Scoring Predictions at Extreme Quantiles	
}


\author{Axel Gandy        \and
        Kaushik Jana \and Almut E. D. Veraart %etc.
}


\institute{Axel Gandy \at
              Department of Mathematics \\ Imperial College London, London, SW7 2AZ, U.K. \\
              \email{a.gandy@imperial.ac.uk}           %  \\
           \and
Kaushik Jana (corresponding author)\at
Department of Mathematics \\ Imperial College London, London, SW7 2AZ, U.K. \\
\email{k.jana@imperial.ac.uk}           %  \\
\and
Almut E. D. Veraart \at
Department of Mathematics \\ Imperial College London, London, SW7 2AZ, U.K. \\
\email{a.veraart@imperial.ac.uk}           %  \\
}

\date{Received: date / Accepted: date}

\maketitle

\begin{abstract}
Prediction of quantiles at extreme tails is of interest in numerous applications. Extreme value modelling provides various competing predictors for this point prediction problem. A common method of assessment of a set of competing predictors is to evaluate their predictive performance in a given situation. However, due to the extreme nature of this inference problem, it can be possible that the predicted quantiles are not seen in the historical records, particularly when the sample size is small. This situation poses a problem to the validation of the prediction with its realisation. In this article, we propose two non-parametric scoring approaches to assess extreme quantile  prediction mechanisms. The proposed assessment methods are based on predicting a sequence of equally extreme quantiles on different parts of the data. We then use the quantile scoring function to evaluate the competing predictors. The performance of the scoring methods is compared with the conventional scoring method and the superiority of the former methods are demonstrated in a simulation study. The methods are then applied to reanalyse cyber Netflow data from Los Alamos National Laboratory and daily precipitation data at a station in California available from Global Historical Climatology Network.

\end{abstract}

\section{Introduction}\label{sec1}
Prediction of extreme quantiles or equivalently, so-called return levels, is an important problem in various applications of extreme value analysis including but not limited to meteorology, hydrology, climatology and finance. The main task of inference in many of these prediction problems involves the computation of probabilities of yet unobserved rare events that are not seen in the historical records. Specifically, the problem may involve the estimation of a high quantile, say at level $p$, based on a sample of size $n$, where $p$ is very close to 1 so that $n(1-p)$ is small. Examples include the prediction of a quantile of the  distribution of precipitation which could be realized once in every 200 years based on 50 to 100 years of rainfall data, or the prediction of high financial loss based on few years of data.  
  
Extreme value theory (EVT) provides tools to predict extreme events, based on the assumption that the underlying distribution of the normalized random variable resides in the domain of attraction of an extreme value distribution \citep{fisher_tippett_1928}. Thus extreme quantiles can be predicted by estimating the parameters of the specific extreme value distribution and the normalizing constants. Detailed reviews on different models and methods of estimation in this context can be found in \cite{Embrechts1997}, and \cite{Coles_2001}, among others. Many distinct models could be applicable for predicting an extreme quantile in a given situation, thereby calling for a comparative assessment of their predictive performance. 

{\color{black}To our knowledge, relatively little work has been done on the evaluation of quantile predictor (point forecast mechanism) in the context of extreme events.} \cite{Friederichs} considered probabilistic forecasts and derived the continuous rank probability score (CRPS) for two common extreme value distributions to assess forecasts for predictive distribution of the peak wind prediction. This method was also used in \cite{bentzien_friederichs_2014} and \cite{scheuerer2015} for precipitation forecast ensembles through the assumption of a parametric form for the prediction distribution. \cite{lerch2017} proposed a weighted CRPS for a probabilistic forecast by imposing weights on the tail of the distribution with emphasis on extreme events. In the case of point prediction, specifically for evaluating quantile prediction, the standard tool to use is the quantile score which is a strictly proper and consistent scoring rule for quantile functional \citep{Gneiting2007}.
{\color{black} In recent work, \cite{brehmer2019} demonstrate that even a proper scoring rule can not distinguish tail properties in a very strong sense. %This implies that a straightforward use of a scoring rule may not be suitable to separate tail quantile forecasts.
} % In \cite{taillardat2019extreme}, CRPS is considered as a random variable, extreme value distribution is used as a model and method is used: explanation is needed.}

{\color{black} Another challenge of assessment of a very high (or a very low) quantile estimate} is that, due to the sparseness of the data at extreme tails, the estimate may fall outside the range of the observed data, therefore, leaving no practical information about the estimate to validate. This instance also implies that the quantile score is optimized trivially at a specific estimate (see equation \eqref{increasing} of Section \ref{sec2}).

To address this problem, we propose to predict a new set of equally extreme quantiles using different subsets of the data set to compare various quantile prediction methods. The new quantile points are chosen such that their predictions are observed in the rest of the data (test sample). Each of the competing predictors is thus used to estimate the series of quantiles on different parts of the data and validated in the rest of the data. This method then uses a combined quantile scoring function and cross-validation to assess different competing extreme quantile predictors.

We propose two approaches. The first one uses a smaller part of the data to predict a new set of quantiles and validate the predictions using the remaining {\color{black}larger} part of the data. In the second approach, the quantiles are predicted on a more substantial part of data and validated in the smaller part of the data. 
%\textcolor{red}{Need to mention that the methods are based on sample splitting and cross validation. The description in on one run of the cross validation. }

We examine the performance of the two methods under different data generating processes through a simulation study and compare their performance with the quantile scoring method. The proposed two methods are more efficient than the quantile scoring method when the GPD model is reasonably well fitted in different parts of the data. We also observe that the first approach, which uses a larger part of data for validation of the competing predictor performs more favourably than the second approach.

The outline of the paper is as follows. In Section~\ref{sec2}, we present two different methods of assessment of the extreme quantile predictors, describe different choices of the tuning parameters, and list a few examples of the extreme quantile predictor. We present a simulation study in Section~\ref{simulation} to demonstrate the performance of the optimum scoring predictors under different data-generating mechanisms. In Section~\ref{data_analysis}, we apply the proposed methodologies to two different data sets: the LANL Netflow data and the precipitation data corresponding to one station of the Global Historical Climatology Network data. We conclude with some remarks in Section~\ref{sec5}.

\section{Methodology}\label{sec2}
Consider identically distributed, continuous random variables $Y, Y_1,\ldots,Y_{n}\in \mathbb{R}$, for $n \in \mathbb{N}$. We write ${\bf Y}=(Y_1,\ldots,Y_{n})$ for the corresponding random sample of size $n$ and  ${\bf y}=(y_1,\ldots,y_{n})$ for a realisation.

Let $p\in (0,1)$. Recall that the $p$-th  quantile of $Y$ is defined as  $$q_Y(p)=\inf\{y: P(Y\leq y)\geq p\}.$$  Let $Q$ denote a quantile predictor function, such that 
 $(p,{\bf Y})\mapsto Q(p,{\bf Y})$ is a predictor of $q_Y(p)$  and $Q(p,{\bf y})$ is its empirical counterpart.

Suppose we wish to estimate the quantile $q_Y(p^0)$ for $p^{0}\in(0,1)$ close to 1 and there are $m\in \mathbb{N}$ competing quantile predictors denoted   by $Q_i$ for $i=1,\ldots, m$. Let 
$$\mathbb{Q}=\{Q_1,\ldots, Q_m\}$$
denote the set of these $m$ competing predictors.

Evaluation of point predictors is typically done by means of a scoring function $s(a,b)$ that assigns a numerical score when the point forecast $a$ is issued and the observation $b$ is realized \citep{Gneiting2011}. For evaluating a point predictor consistently, it is recommended to use a strictly proper and consistent scoring rule for the functional of interest. The quantile score is such a rule for assessing quantile predictors at given probability levels and used in various applications \citep{Bentzien2014}.

In the context of evaluating quantile forecasts, one typically uses the check loss function, see \citep{koenker1984}, defined as
$$\rho_{p}(a,b)=(a-b)(p-I(a<b)), \qquad \mathrm{ for }\; a, b \in \mathbb{R}.$$ 

We will now introduce our notation for the corresponding quantile scores we are going to consider. Let ${\bf x}^{(1)}$ and ${\bf x}^{(2)}$ denote $n_1$- and, respectively,  $n_2$-dimensional realisations of subsamples of ${\bf Y}$. Then we define by
\begin{align}\label{eq:ScoreDef}
    S(Q(\cdot,\cdot), p, {\bf x}^{(1)}, {\bf x}^{(2)})
    =\frac{1}{n_2}\sum_{i=1}^{n_2}\rho_p(Q(p,{\bf x}^{(1)}),x_i^{(2)})
\end{align}
the average quantile score for the predictor $Q(p,\cdot)$ based on the empirical test (sub-) sample ${\bf x}^{(1)}$ and the empirical   validation (sub-) sample ${\bf x}^{(2)}$.

In the case when ${\bf x}^{(1)}={\bf x}^{(2)}={\bf y}$, equation \eqref{eq:ScoreDef} simplifies to 
\begin{align}\label{eq:ScoreSimple}
    S(Q(\cdot,\cdot), p, {\bf y}):=S(Q(\cdot,\cdot), p, {\bf y}, {\bf y})
    =\frac{1}{n}\sum_{i=1}^{n}\rho_p(Q(p,{\bf y}),y_i).
\end{align}

Given the set $\mathbb{Q}$ of competing predictors, the optimum quantile predictor $\widehat {Q}_{qs}$ is defined as the minimizer of the average quantile score via
\begin{equation}\label{opt1}
\widehat {Q}_{qs}=
\widehat {Q}_{qs}(\cdot, \cdot)=\mathop{\arg\min}_{Q\in \mathbb{Q}}S(Q(\cdot,\cdot), p^0,{\bf y}).
\end{equation} 
 Then 
 $\widehat q_{qs}(p^0)=\widehat {Q}_{qs} (p^0, {\bf y})$  denotes estimated $p^0$-quantile using  the predictor $\widehat Q_{qs}$ and the data ${\bf y}$.

Predicting quantiles from very extreme tails is a difficult task due to the sparseness of the data in the tails of the distribution, especially for heavy-tailed distributions, see \citep{wang2012}. This fact also makes it challenging to evaluate the predictions, as they may not be realized inside the range of the observed data. 

In the situation, where all predictors in $\mathbb{Q}$ predict a quantile larger than $\max\{y_1,\ldots,y_n\}$, the optimal quantile prediction is the smallest prediction from $\mathbb{Q}$. Indeed, for every $Q(p, {\bf y})\in \mathbb{Q}$, equation \eqref{eq:ScoreSimple} reduces to 
\begin{equation}\label{increasing}
S(Q(p,\cdot),{\bf y})=\frac{p}{n}\sum_{i=1}^{n}\bigl(Q(p, {\bf y})-y_i\bigr).
\end{equation} 
This may not be a reasonable answer when assessing the prediction of very large quantiles. 
 
This shortcoming motivated us to consider an improved scoring method for assessing predictors for high quantiles. The developed method can be easily adapted to the extreme lower quantiles with $p$ close to 0. 

More specifically, we propose two methods for evaluating the performance of the predictors of $\mathbb{Q}$ by training and assessing them on different parts of the data set. Both methods employ a $k$-fold cross-validation, where 
Method 1 uses a small training and large validation sample, whereas Method 2 uses a large training and a small validation sample.
% for predicting a new set of quantiles. 

\subsection{Main idea}

Suppose we want to predict the $p^{0}$th quantile for  $p^0$ close to 1. 
Let us consider the problem of predicting the $p^{c}$th quantile where, $0<p^{c}<p^0<1$, based on a sub-sample (training sample)  of size $n^{c}$ of the original sample of size $n$.
%We denote the training sample by ${\bf t}$.

We then propose to choose $p^{c}$ and $n^{c}$ such that estimating the $p^{c}$th quantile based on a sub-sample of size $n^{c}$ is \emph{equally extreme} as  estimating the $p^0$th quantile based on the original sample of size $n$. By equally extreme, we mean, the expected number of exceedances above the estimated $p^{c}$th and $p^0$th quantile are the same in training and original sample, respectively. This condition implies that  
\begin{equation}\label{cond1}
n^{c}(1-p^{c})=n(1-p^0).
\end{equation}
Now we have one equation and two unknowns ($n^{c}$ and $p^{c}$). In order to be able to 
determine $n^{c}$ and $p^{c}$ uniquely, we impose an additional condition:

Let $\alpha>0$ denote the \emph{
expected number of observations exceeding the $p^{c}$th quantile in the remainder of the sample}, i.e.~in the validation sample of size $n-n^c$.
We assume that $\alpha$ is fixed a priori, hence we obtain the condition that, 
 for a {\it given} $\alpha$:
\begin{eqnarray}\label{cond2}
(n-n^{c})(1-p^{c})=\alpha. 
\end{eqnarray} 
The system of the two equations \eqref{cond1} and \eqref{cond2} has  a unique solution for $n^{c}$ and $p^{c}$ given by
\begin{eqnarray}
n^{c}&=&\frac{n}{1+\frac{\alpha}{n(1-p^0)}},\label{cond-n}\\
p^{c}&=&p^0-\frac{\alpha}{n}\label{cond-p}.
\end{eqnarray}
In practice, the choice of $\alpha$ will determine the balance between the number of exceedances in the training and the validation sample. 
A smaller (or larger) value of $\alpha$ leads to a new quantile ($p^{c}$th) closer to (or further away from)  the target quantile ($p^0$th) and leaves fewer (more) observations for the evaluation of the predictions.

\subsubsection{Using $k$-fold cross-validation}
%We propose to  two approaches for implementing our idea in a $k$-fold cross-validation framework. 

We will now describe how our idea can be implemented in 
 a $k$-fold cross-validation framework.
 
Suppose $n$ is given and $p^0$ and $\alpha$ have been fixed,  and $n^c$ and $p^c$ have been determined using equations \eqref{cond-n} and \eqref{cond-p}, respectively.

We split the original sample ${\bf y}$ into $k$ sub-samples of size $\frac{n}{k}$ denoted by ${\bf y}^{(i)}$, for $i=1, \ldots, k$. Also, ${\bf y}^{\setminus(i)}$ denotes the sub-sample of ${\bf y}$ of size $\frac{(k-1)}{k}n$, %$n-n/k=\frac{(k-1)}{k}n$, 
which excludes all the observations contained in ${\bf y}^{(i)}$.

We have to find $k$ for fixed $n, p^0, \alpha$. In order to do this, we investigate two routes:
\begin{itemize}
   \item Method 1: Small training sample (one fold) and large validation sample (k-1 folds).
   \item Method 2: Large training sample (k-1 folds) and small validation sample (one fold).
\end{itemize}

We note that the choice of $\alpha$ determines how much importance is given to the estimation versus the validation of the extremes. A relatively high value of $\alpha$  leads to more exceedances available for validation, but less for estimation and vice versa.

We note that in the following we will be treating  $\alpha$ as our tuning parameter and derive an expression for the number of folds $k$ in terms of $\alpha$. Equivalently, one could fix the number of folds $k$ first and derive the corresponding expression for $\alpha$.
\subsection{Method 1: Small training sample and large validation sample}\label{proposed-method1}

We propose an unconventional version of the $k$-fold cross-validation, where 
 at a time, one sub-sample ${\bf y}^{(i)}$ is used as the training sample and the collection of the remaining  $k-1$ sub-samples given by ${\bf y}^{\setminus(i)}$ is used as the validation sample. Thus each data point is used once for training and  $k-1$ times for validation purposes. % in the entire process. 
(In a conventional $k$-fold cross-validation set-up, one would use a collection of $k-1$ folds for training and one fold  for validation.)

I.e.~we set the size of the training sample as $n^c=n/k$, which results in
\begin{align*}
 k=\frac{n}{n^c}=1+\frac{\alpha}{n(1-p^0)} \Longleftrightarrow 
 \alpha =   n(1-p^0)(k-1).
\end{align*}
Here we see that we can either fix $\alpha$ and solve for $k$ or vice versa.
In the former case, we would typically set
$k=\lfloor 1+\frac{\alpha}{n(1-p^0)} \rfloor$,
to ensure that $k$ is integer-valued.
Also, recall that $p^c=p^0-\frac{\alpha}{n}$. 

%In our setting, This implies that if $\alpha$ number of observations exceeding the predicted quantile would have been observed in the original sample, then $\alpha(k-1)$ exceedances (above the predicted number) can be observed in the test sample in the whole process. 

We can then define the score based on the $k$-fold cross-validation as
\begin{equation}\label{score2_fixedalpha}
SCV^{(1)}(Q, p^0, \alpha, {\bf y})=
	 \frac{1}{k}\sum_{j=1}^{k}
	S(Q(\cdot,\cdot), p^c, {\bf y}^{(j)}, {\bf y}^{\setminus(j)}).
\end{equation}
	%where we recall that $p^c=p^0-\frac{\alpha}{n}$ and $k=[n/n^c]=[1+\frac{\alpha}{n(1-p^0)}]$.
	
	Since the tuning parameter $\alpha$ clearly affects the score, we propose to define a more general score in the next subsection, which takes a variety of choices of $\alpha$ into account.
	
\subsubsection{Using a variety of tuning parameters $\alpha$}
Consider $l\in \mathbb{N}$ distinct values of the tuning parameter $\alpha$ as $\alpha_1, \alpha_2, \ldots,\alpha_l$ and set ${\boldsymbol \alpha}=(\alpha_1, \alpha_2, \ldots,\alpha_l)$. For each choice $\alpha_i$ for $i=1,\ldots, l$, we derive the corresponding  $n^c_i$ and $p^c_i$ via the equations \eqref{cond-n} and \eqref{cond-p}. I.e.
\begin{align*}
n^{c}_i=\frac{n}{1+\frac{\alpha_i}{n(1-p^0)}},&&
p^{c}_i=p^0-\frac{\alpha_i}{n}.
\end{align*}
Then we set $k_i=\lfloor n/n_i^c\rfloor=\lfloor 1+\frac{\alpha_i}{n(1-p^0)}\rfloor$.

For a  predictor $Q\in \mathbb{Q}$, we define a combined score for prediction of the $p^0$th quantile by
\begin{equation}\label{score2_fixedalphavector}
SCV^{(1)}(Q, p^0, {\boldsymbol \alpha}, {\bf y})=\frac{1}{l}\sum_{i=1}^l
	 \frac{1}{k_i}\sum_{j=1}^{k_i}
	S(Q(\cdot,\cdot), p_i^c, {\bf y}^{(j)}_i, {\bf y}^{\setminus(j)}_i), 
	\end{equation}
where   ${\bf y}^{(j)}_i$ is the $j$-th sample fold of size $n_i^c=n/k_i$ of the original sample ${\bf y}$,
and 
${\bf y}^{\setminus(j)}_i$ denotes the collection of the $k_{i}-1$ sub-samples of ${\bf y}$ excluding ${\bf y}^{(j)}_i$.

	This combined score function pools information from assessing a given predictor at multiple and similar extreme trial quantile levels. 
	
	\begin{remark}
	We note that we have assigned equal weights to the cross-validation scores corresponding to different choices of $\alpha$. It will be interesting to explore in more detail whether alternative weights should be considered when combining the individual scores.
	\end{remark}
	
	We define the  optimal predictor  as the minimizer of the combined quantile score   
\begin{equation}\label{opt2}
\widehat{Q}_1
=\widehat{Q}_1(\cdot, \cdot)=\mathop{\arg\min}_{Q\in \mathbb{Q}}SCV^{(1)}(Q, p^0, {\boldsymbol \alpha}, {\bf y}).
\end{equation}
Note that the dependence of $\widehat Q_1$ on the choice of the tuning parameter ${\boldsymbol \alpha}$ is not reflected in the notation to simplify the exposition.

 Then 
 $\widehat q_{1}(p^0)=\widehat {Q}_{1} (p^0, {\bf y})$  denotes estimated $p^0$-quantile using  the predictor $\widehat Q_{1}$ and the data ${\bf y}$.

\subsection{ Method 2: Large training sample and small validation sample}\label{proposed-method2}
%\subsubsection{\color{black} Large training sample and small validation sample}

Alternatively, one could use a traditional $k$-fold cross-validation set up to compare the competing predictors, where $(k-1)$ sub-samples are used for training and just one sub-sample for validation. 
Hence we have
\begin{align} \label{cond-n-method2}
    n^c=(k-1)\frac{n}{k}. 
\end{align}
The number of folds $k$ needs to be determined depending on the fixed parameters $n, p^0, \alpha$. Using equations \eqref{cond-n} and \eqref{cond-n-method2}, we solve for $k$ and get
\begin{align*}
    k=\frac{n(1-p^0)}{\alpha}+1 \Longleftrightarrow \alpha=n(1-p^0)(k-1)^{-1}.
\end{align*}
Hence, we set 
\begin{align}\label{k-Method2}
k=\left\lfloor \frac{n(1-p^0)}{\alpha}+1\right\rfloor.
\end{align}
As before, we also have that
\begin{align}\label{cond-p-method2}
    p^c=p^0-\frac{\alpha}{n}.
\end{align}

As in Method 1, we can then define the score based on the $k$-fold cross-validation as
\begin{equation}\label{score2_fixedalpha_method2}
SCV^{(2)}(Q, p^0, \alpha, {\bf y})=
	 \frac{1}{k}\sum_{j=1}^{k}
	S(Q(\cdot,\cdot), p^c, {\bf y}^{\setminus (j)}, {\bf y}^{(j)}).
	\end{equation}

\subsubsection{Using a variety of tuning parameters $\alpha$}
	As for Method 1, we can also consider  a variety of choices of the tuning parameter  $\alpha$.
	As before, consider $l\in \mathbb{N}$ distinct values of the tuning parameter $\alpha$ as $\alpha_1, \alpha_2, \ldots,\alpha_l$ and set ${\boldsymbol \alpha}=(\alpha_1, \alpha_2, \ldots,\alpha_l)$. For each choice $\alpha_i$ for $i=1,\ldots, l$, we derive the corresponding  $n^c_i$ and $p^c_i$ via the equations \eqref{cond-n-method2} and \eqref{cond-p-method2}. I.e.
\begin{align*}
n^{c}_i=n^c=(k-1)\frac{n}{k},&&
p^{c}_i=p^0-\frac{\alpha_i}{n}.
\end{align*}
%Then we set  $k_i=\left[\frac{n(1-p^0)}{\alpha_i}+1\right]$.

For a  predictor $Q\in \mathbb{Q}$, we define a combined score for prediction of the $p^0$th quantile by
\begin{equation}\label{score2_fixedalphavector-method2}
SCV^{(2)}(Q, p^0, {\boldsymbol \alpha}, {\bf y})=\frac{1}{l}\sum_{i=1}^l
	 \frac{1}{k_i}\sum_{j=1}^{k_i}
	S(Q(\cdot,\cdot), p_i^c, {\bf y}^{\setminus(j)}_i, {\bf y}^{(j)}_i), 
	\end{equation}
%where $k_i=[1+\frac{\alpha_i}{n(1-p^0)}]$ and % with $[a]$ is the greatest integer less or equal to $a$, 
%$p_i^{c}=p^0-\frac{\alpha_i}{n}$. 
where   ${\bf y}^{(j)}_i$ is the $j$-th sample fold of size $n/k_i$ of the original sample ${\bf y}$, %, where $n^c_1=\ldots=n^c_k=n^{c}$, 
and 
${\bf y}^{\setminus(j)}_i$ denotes the collection of the $k_{i}-1$ sub-samples of ${\bf y}$ excluding ${\bf y}^{(j)}_i$.

	We define the  optimal predictor  as the minimizer of the combined quantile score   
\begin{equation}\label{opt2-Method2}
\widehat{Q}_2=\widehat{Q}_2(\cdot, \cdot)=\mathop{\arg\min}_{Q\in \mathbb{Q}}SCV^{(2)}(Q, p^0,  {\boldsymbol \alpha}, {\bf y}).
\end{equation}
Note again that the dependence of $\widehat Q_2$ on the choice of the tuning parameter ${\boldsymbol \alpha}$ is not reflected in the notation.

Then 
 $\widehat q_{2}(p^0)=\widehat {Q}_{2} (p^0, {\bf y})$  denotes estimated $p^0$-quantile using  the predictor $\widehat Q_{2}$ and the data ${\bf y}$.
\subsection{Summary of the two methods and choice of $\alpha$}\label{alpha_choice}
We summarise the set-up of Methods 1 and 2 in Table \ref{tab:SummaryMethods}. 
 \begin{table}[h]\centering
\ra{1.3}
%\begin{tabular}{@{}ccc@{}}\toprule
{\footnotesize \begin{tabular}{cccc}\toprule
    Description & Requirement & Method 1 & Method 2  \\ \midrule
    Training sample size $n^c$ & $\frac{n}{1+\frac{\alpha}{n(1-p^0)}}$&$\frac{n}{k}$& $(k-1)\frac{n}{k}$\\
    Validation sample size $n-n^c$ & $\frac{\alpha n}{n(1-p^0)+\alpha}$& $(k-1)\frac{n}{k}$&$\frac{n}{k}$\\
    Number of folds $k$ & &$%\lfloor \frac{n}{n^c}\rfloor=
    \lfloor 1+ \frac{\alpha}{n(1-p^0)}\rfloor$
    & $%\lfloor \frac{n}{n-n^c}\rfloor =
    \lfloor 1+\frac{n(1-p^0)}{\alpha}\rfloor$\\
    $p^c$& $p^0-\frac{\alpha}{n}$ & & \\
    $(n-n^c)(1-p^c)$& $\alpha$ 
    & %$(k-1) \frac{n}{k}(1-p^c)$
    $n(1-p^0)(k-1)$
    &%$\frac{n}{k}(1-p^c)$
    $n(1-p^0)(k-1)^{-1}$\\
     $n(1-p^c)$&$n(1-p^0)+\alpha$ &
     $n(1-p^0)k$&$n(1-p^0)\frac{k}{k-1}$   \\
    % \midrule
    %  \multicolumn{3}{c}{Case of multiple $\alpha$s: $\boldsymbol{\alpha}=(\alpha_1,\ldots,\alpha_l)$}\\
\bottomrule
\end{tabular}
}
\caption{Overview of the two methods. Recall that $\alpha$ denotes the expected number of exceedances above the $p^c$th quantile in the validation sample. The quantity $n(1-p^c)$ represents the expected number of exceedances above the $p^c$ quantile in the entire sample.\label{tab:SummaryMethods}}
\end{table}
%subsubsection{Choice of $\alpha$}\label{alpha_choice}
Recall that the tuning parameter  $\alpha$ specifies the average number of observations in the validation data that exceeds the $p^c$- quantile estimated in the training sample. The performance of the predictors are evaluated in the validation sample and the best predictor is the minimizer of the average quantile score. % given in the parenthesis of \eqref{score2}.
Therefore, the best predictor is the optimum quantile estimate (according to the quantile score) in the validation sample among all quantile predictors under consideration. But for a given validation sample of fixed size, it is only possible to get a good classical optimizer when there are at least few observations larger than the desired quantile. Hence, the tuning parameter $\alpha$ has to be chosen accordingly. 

We note that suitable ranges of $\alpha$ will differ depending on which of the two methods is used. While Method 1 can allow for larger values of $\alpha$ (in the sense of $\alpha>1$), we would expect that in Method 2, we should choose  $\alpha<1$ to reflect the fact that, for extreme quantiles at levels  $p^c<p^0$, we would not expect to see exceedances in a relatively small validation sample based on only one fold in the cross-validation. 

In our simulation study and empirical work, we consider the choice that $p^0=1-\frac{1}{2n}$, in which case the target quantile is so extreme that we might not even see a single observation above that threshold. We summarise the two methods for this particular case in Table \ref{tab:SummaryMethods-Ex}.

\begin{table}[h]\centering
\ra{1.3}
%\begin{tabular}{@{}ccc@{}}\toprule
{\footnotesize \begin{tabular}{cccc}\toprule
    Description & Requirement & Method 1 & Method 2  \\ \midrule
    Training sample size $n^c$ & $\frac{n}{1+2\alpha}$&$\frac{n}{k}$& $(k-1)\frac{n}{k}$\\
    Validation sample size $n-n^c$ & $\frac{2\alpha n}{1+2\alpha}$& $(k-1)\frac{n}{k}$&$\frac{n}{k}$\\
    Number of folds $k$ & &$%\lfloor \frac{n}{n^c}\rfloor=
    \lfloor 1+2\alpha\rfloor$
    & $%\lfloor \frac{n}{n-n^c}\rfloor =
    \lfloor 1+\frac{1}{2\alpha}\rfloor$\\
    $p^c$& $1-\frac{(1+2\alpha)}{2n}$ & & \\
    $(n-n^c)(1-p^c)$& $\alpha$ 
    & %$(k-1) \frac{n}{k}(1-p^c)$
    $\frac{1}{2}(k-1)$
    &%$\frac{n}{k}(1-p^c)$
    $\frac{1}{2}(k-1)^{-1}$\\
     $n(1-p^c)$&$\frac{1}{2}+\alpha$ &
     $\frac{1}{2}k$&$\frac{1}{2}\frac{k}{k-1}$   \\
    % \midrule
    %  \multicolumn{3}{c}{Case of multiple $\alpha$s: $\boldsymbol{\alpha}=(\alpha_1,\ldots,\alpha_l)$}\\
\bottomrule
\end{tabular}
}
\caption{Special case of Table \ref{tab:SummaryMethods} when when $p^0=1-\frac{1}{2n}$.
\label{tab:SummaryMethods-Ex}
}
\end{table}

For Method 1, motivated by work by \cite{wang2012} and \cite{Velthoen2019}, we choose $\alpha \in \mathbb{N}\cap [1, n^{1/4}]$. A restriction to integer-valued $\alpha$s is not necessary but aids our interpretation.
We then determine the corresponding values of $k$ based on the choice of $\alpha$.

In order to make the comparison between Methods 1 and 2 easier, we then take the values of $k$ derived in Method 1 and convert them to the corresponding values of $\alpha$ in Method 2. For $n=7500$, this procedure is summarised in Table \ref{tab:Choicealpha}.

\begin{table}[h]\centering
\ra{1.3}
%\begin{tabular}{@{}ccc@{}}\toprule
{\footnotesize \begin{tabular}{cccccc}\toprule
   &k &  3& 5 & 9& 17 \\ \midrule
 \multirow{2}{*}{ Method 1} &$\alpha$  & 1&2&4&8\\
  &$p^c=1-\frac{(1+2\alpha)}{n}$  & $1-\frac{3}{n}$& $1-\frac{5}{n}$ & $1-\frac{9}{n}$&$1-\frac{17}{n}$\\ \midrule
  \multirow{2}{*}{ Method 2} &$\alpha$ &$\frac{1}{2}$& $\frac{1}{8}$&$\frac{1}{16}$&$\frac{1}{32}$\\
  & $p^c=1-\frac{(1+2\alpha)}{2n}$ &
  $1-\frac{2}{2n}$ &$1-\frac{5}{8n}$&$1-\frac{9}{16n}$&$1-\frac{17}{32n}$\\
\bottomrule
\end{tabular}
}
\caption{Parameter choices for the simulation study: Consider the example when $p^0=1-\frac{1}{2n}$. 
From the formulas in the table we can read of that 
a choice of $\alpha \in \{1, 2, 4, 8\}$ in Method 1 corresponds to $k\in \{3, 5, 9, 17\}$. Also, choosing $\alpha \in \{1/4, 1/8, 1/16, 1/32\}$ in Method 2 also corresponds to  $k\in \{3, 5, 9, 17\}$.\label{tab:Choicealpha}
}
\end{table}

Related to our choice of multiple $\alpha$s, we note that the use of multiple quantile levels to estimate regression coefficients appeared in the quantile regression context \citep{koenker1984, koenker2004, zou2008} and in extreme quantile estimation \citep{wang2012, Velthoen2019}. 
%Here each of the quantile levels is chosen such that the optimal asymptotic properties of the regression parameter estimates can be achieved. 

\subsection{Examples of extreme quantile predictor}\label{predictor}
The two commonly used models for predicting extreme quantiles are the Generalized Extreme Value Distribution (GEV) for block maxima and the Generalized Pareto Distribution (GPD) for peaks-over-thresholds of the variable of interest.

{\bf GEV model:} The GEV$(\mu,\sigma,\xi)$ model is parametrized by the location parameter $\mu$, the scale parameter $\sigma$ and the shape parameter $\xi$. Its cumulative distribution function is given by  
\begin{equation}\label{gev1}
F_Y(y|\mu,\sigma, \xi)=\begin{cases} \exp \left(-\left(1+\xi  \frac{y-\mu}{\sigma}\right)^{-1/\xi}\right),\;\; \xi\geq 0,\\
\exp \left(-\exp\left(-  \frac{y-\mu}{\sigma}\right)\right),\;\; \xi= 0,\\
\end{cases}
\end{equation}
where $1+\xi[(y-\mu)/\sigma]>0$ for $\xi \neq 0$. 

For a given level $p$ and a realisation ${\bf y}$, the $p$-th quantile of $Y$ can be estimated as
\begin{equation}\label{gev-est}
Q(p, {\bf y})=\begin{cases}\hat\mu+\frac{\hat\sigma}{\xi}\left[1-(-\log p) ^{-\hat\xi}\right],& \text{if } \hat \xi\neq 0,\\\hat\mu-\hat\sigma \log(-\log p), & \text{if } \hat \xi= 0, 
\end{cases}
\end{equation} 
where $\hat{\mu}$, $\hat{\sigma}$ and $\hat{\xi}$ are estimates of ${\mu}$, ${\sigma}$ and ${\xi}$.

{\bf GPD model:} Let us assume that there exists a non-degenerate limiting distribution for appropriately linearly rescaled excesses of a sequence of independently and identically distributed observations $Y_1,\ldots,Y_n$ above a threshold $u$. Then under general regularity conditions, the limiting distribution will be a GPD as $u\rightarrow \infty$ \citep{Pickands_1975}. The GPD is parametrized by a shape parameter $\xi$ and a threshold-dependent scale parameter $\sigma_u$, with cumulative distribution function $F$, 
\begin{equation}\label{gpd1}
F_Y(y|u,\sigma_u, \xi)=\begin{cases} 1-\left[1+\xi \left( \frac{y-u}{\sigma_u}\right)\right]^{-1/\xi},\;\; y \geq u, \xi\geq 0,\\
1-\exp\left[\frac{y-u}{\sigma_u}\right],\;\;\;\; u<y<u-\sigma_u/\xi, \xi<0,
\end{cases}
\end{equation}
and $u\in \mathbb{R}$ and $\sigma_u>0$. 

The GPD model depends on a threshold $u$, which needs to be chosen up front. The choice of the threshold is a crucial issue, as too low of a threshold leads to a bias from model misspecification, and too high of a threshold increases the variance of the estimators: a bias-variance trade-off. Many threshold selection procedures have been proposed in the literature, see  \citep{Davison1990, drees2000,Coles_2001}, among many others.

Once an appropriate threshold has been selected, the parameters can be estimated by the maximum likelihood method and, subsequently, the target high quantile estimate can be extracted by plugging in the estimates of $u$, $\xi$ and $\sigma_u$ in 
\begin{equation}\label{gpd-est}
Q(p, {\bf y})= \begin{cases}u+\frac{\hat\sigma}{\hat\xi}\left[\left(\frac{\hat\zeta_u}{1-p}\right)^{\hat\xi}-1\right],& \text{if } \xi\neq 0,\\u+\hat\sigma \log\left(\frac{\hat\zeta_u}{1-p}\right), & \text{if } \xi= 0,
\end{cases}
\end{equation} 
where $\hat\sigma$, and $\hat\xi$ are the maximum likelihood estimates based on the GPD assumption for the exceedances (\citep{Coles_2001}, Section 4.3.3). Also, $\zeta_u=P(Y>u)$ can be approximated by the ratio $\frac{n_u}{n}$ where $n_u$ is the number of exceedances above the threshold $u$ in the given sample of size $n$. 

%\subsection{GPD threshold selection using the scoring method}Selection of threshold is a crucial issue in fitting a GPD distribution to the exceedances over a  over the threshold. The method proposed in Section~\ref{proposed-method} can be used to select suitable threshold.  

\section{Simulation Study}\label{simulation}
In this section, we carry out a simulation study to examine the performance of the optimum score predictors $\widehat{Q}_1$ and $\widehat{Q}_2$ obtained through the proposed scoring methods and compare them with that of the conventional optimal quantile score predictor $\widehat{Q}_{qs}$ at large quantiles.   

We simulate samples from the random variable $Y$ following the four sets of distribution with densities: 

\begin{itemize}
	\item[(i)]  $f(y)=f_{GPD (10, \;1,\; \xi)} (y)$ with three choices of the shape parameter $\xi$: ($a$) $\xi$= -0.5 , ($b$) $\xi$=0, and ($c$) $\xi$=0.5,
	\item[(ii)] $f(y)=\lambda f_{U(0, \;10)}(y)+ (1-\lambda) f_{GPD (10, \;1, \;0.5)} (y)$, with two different choices of the weight parameter $\lambda$: ($a$) $\lambda$= 0.5  and ($b$) $\lambda$=0.99,
	\item[(iii)] $f(y)=0.5f_{GPD (10, \;1, \;0.1)} (y)+ 0.5 f_{GPD (10, \;1, \;0.5)} (y)$,
	\item[(iv)] $f(y)=f_{Gamma (1, 1, 0.1)} (y)$,
\end{itemize}
where $f_{U(0, 10)}$ denotes the probability density function (pdf) of the Uniform distribution (0, 10) and $f_{GPD(\mu,\; \sigma,\; \xi)}$ denotes the pdf of the GPD with location $\mu$, scale $\sigma$, and shape $\xi$. The three different values of the shape parameter in model-$(i)$ correspond to the three types of the extreme value families: $\xi=-0.5$ (Weibull), $\xi=0$ (Gumbel) and $\xi=0.5$ (Frechet). We choose the standard scale $\sigma=1$ and location $\mu=10$ in each of the cases: $(i)$ to $(iii)$. The model-$(ii)$, is a mixture with mixing probability $\lambda$ such that it produces lower 100$\lambda$ percent samples from $U(0, 10)$ and upper 100(1-$\lambda$) percent samples from the GPD. We choose $\lambda=0.5$ and $\lambda=0.99$ in model-$(ii)$. The model-$(iii)$ is obtained by equally mixing two GPDs with different shape parameters. Model-$(iv)$ is the Gamma distribution with rate, scale and shape parameter taking values as 1, 1 and 0.1, respectively. 

For each of the simulated datasets, we apply the conventional method $\widehat{Q}_{qs}$ and the proposed scoring methods $\widehat{Q}_1$ and $\widehat{Q}_2$ to evaluate and select the optimum score predictors of the quantile at the level $p^0=(1-\frac{1}{2n}$) based on a sample of size $n$. We take sample size $n=7500$, this value of $n$ can be thought of as 50 years of daily observations (150 days each season) in the precipitation rate data analyzed in Section \ref{ghcn}.

As a competing set of predictors for the target quantile level $p^0$, we consider the GPD-model-based predictors given in \eqref{gpd-est} estimated by the maximum likelihood method. For selecting the threshold in the GPD model, we consider the ``rules-of-thumb" approach \citep{Ferreira2003}, which uses a fixed upper fraction of data to fit the GPD. We consider two sets of thresholds. The first set is based on ten upper order statistics: 150, 125, 100, 75, 50, 40, 30, 20, 10 and 3, and refer to this set as $A$. Here the largest order statistics corresponds to the lowest GPD threshold. We label the predictors $Q_i$ where, for the index $i$,  we use numbers starting from 1 to  10, such that, 1 and 10 correspond to the predictors with the highest and lowest value of the upper order statistics, respectively.

Note that, each of the members of the set $A$ uses an equal number of observations fitted to the GPD in both the training and original prediction. We then consider another set of thresholds, determined by sample percentiles with probability levels: 0.98, 0.9833, 0.9867, 0.99, 0.993, 0.995, 0.996, 0.9973, 0.9987, and 0.9996, and refer this set as $B$. These probability levels are chosen such that the number of exceedances is approximately the same for $A$ and $B$ for the original sample of size $n=7500$. Note that for each of the members of $B$, the number of samples fitted to the GPD in the original prediction is larger than those in the training stage (\textcolor{black}{as the values of the thresholds are different for different sample sizes}).  For notational convenience, as in set $A$, we index the members of $B$ by the numbers starting from 11 to 20. \textcolor{black}{We also added the sample quantile at level $p^0$ as a predictor in the competing set and denote it as 0-th estimate. Here, out of sample quantiles are estimated by the maximum of the data set.}

For choosing the values of the tuning parameter $\alpha$ for $\widehat{Q}_1$ and $\widehat{Q}_2$, we use the strategy
summarised in Table  \ref{tab:Choicealpha}, resulting in ${\boldsymbol \alpha}= (1, 2, 4, 8)$ (where  $[n^{1/4}]\approx 9$). We found that the use of additional values of $\alpha$ does not change the performance of the proposed methods noticeably.

We generate $L= 1000$ replicates each of size $n=7500$ from the models $(i)$--$(iv)$. For each of the simulated dataset, we use three methods, $\widehat{Q}_{qs}$, $\widehat{Q}_1$ and $\widehat{Q}_2$ to find the optimum quantile score predictors chosen from the competing predictor-set of $p^0$th quantile, and for each of the three methods, we compute the root mean squared error (RMSE) across $L$ iterations as 
\begin{equation}\label{rmse0}
RMSE=\sqrt{\frac{1}{L}\sum_{i=1}^L (\widehat q^{(i)}_{\cdot}(p^0)-q_Y(p^0))^2},
\end{equation}
where $\widehat q^{(i)}_{\cdot}(p^0)$ is the optimal $p^0$-quantile prediction in the $i$th iteration, which is based on any of the three predictors: $\widehat{Q}_{qs}$, $\widehat{Q}_1$ and $\widehat{Q}_2$. For instance, if $\widehat Q_1$ is the optimal predictor in the $i$th iteration, then 
\begin{align*}
   \widehat q^{(i)}_{\cdot}(p^0)= \widehat q^{(i)}_{1}(p^0)=
   \widehat Q_1(p^0,{\bf y}).
\end{align*}

%{\bf \color{red} Dear Kaushik, can you please double-check the notation and numerical values from here to make sure that they align with the new set-up, this applies in particular to Method 2. Many thanks!} 

Table~\ref{tab-rmse001} shows the RMSE of the three optimum predictors $\widehat{Q}_{qs}$, $\widehat{Q}_1$ and $\widehat{Q}_2$ chosen from the sets of the predictors $A$, $B$ and union of the two sets, $A\cup B$. The first proposed approach $\widehat{Q}_1$ outperforms the conventional method $\widehat{Q}_{qs}$ in most of the cases. The only exception is model-$(ii): (b)$, where the optimum predictor $\widehat{Q}_1$ produces a higher RMSE than $\widehat{Q}_{qs}$. The above data generating model consists of  fewer GPD components, on average only 75 GPD observations in the whole sample. This returns roughly four GPD samples in the training data of size $n_i^{c}=441$ for a given $\alpha_4=8$. An inadequate sample from the GPD implies that most of the competing GPD based predictors in $A\cup B$ are poorly fitted, thus resulting in a high bias and a high RMSE. An increase of the GPD component in $(ii): (a)$ (50 per cent data from GPD) returns a lower RMSE for $\widehat{Q}_1$ compared to $\widehat{Q}_{qs}$. It is also found that the presence of only 5 per cent of the GPD component in model-$(ii)$ (for $n=7500$ and $\alpha=8$ it provides at least 20 GPD random sample in the training sample) enables the competing predictors to fit the GPD reasonably well. In this scenario, $\widehat{Q}_1$ outperforms the other two methods (results are not reported here).

Note that, the conventional scoring method $\widehat{Q}_{qs}$ uses the whole sample of size 7500 to train the predictor set. Therefore, in the case of model-$(ii), b$, all the competing predictors are trained on at least 75 GPD sample points (1 per cent of the total sample). Thus they are reasonably well fitted to the GPD, and $\widehat{Q}_{qs}$ outperforms $\widehat{Q}_1$ (row 4 of table~\ref{tab-rmse001}). It can be concluded that if the competing GPD based predictors are reasonably well fitted in the training sample, then the proposed method $\widehat{Q}_1$ performs better than $\widehat{Q}_{qs}$. The latter estimator is more efficient than the former when the data contains only a few extreme components.

In contrast to $\widehat{Q}_1$, the second approach $\widehat{Q}_2$ uses most of the data to train the competing predictors. For example, in model $(ii):(b)$ the method uses a training sample of size 5000 (with $k=3$ in \eqref{cond-n-method2}) consisting of 50 GPD sample. The presence of a sufficient number of GPD samples enables all the predictors to be well fitted to the GPD, and the optimum predictor $\widehat{Q}_2$ produces a smaller error than $\widehat{Q}_1$. Table~\ref{tab-rmse001} also shows that $\widehat{Q}_2$ outperforms the classical optimum score predictor $\widehat{Q}_{qs}$ in all the cases. However, it is evident that, in general, $\widehat{Q}_1$ is more efficient than $\widehat{Q}_2$.  The superiority of $\widehat{Q}_2$ over $\widehat{Q}_1$ in model-$(ii) :(b)$, could be because of a  better estimation of the competing quantile predictors (based on a larger training sample of size 6750) and at the same time having a reasonable amount of exceedances in the cross-validation (for example $\alpha=8$).

\begin{figure}
	\centering
	\includegraphics[height=2.75in]{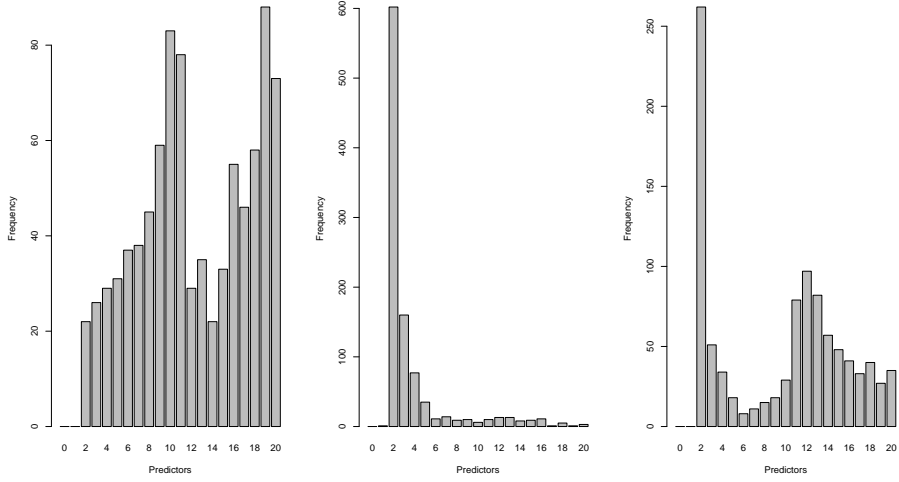}
	\caption[]{Frequency distributions of optimum predictors through the scoring methods $\widehat{Q}_{qs}$, $\widehat{Q}_1$ and $\widehat{Q}_2$ plotted in the left, middle and right panel, respectively. The data are simulated from $(ii):(a)$. The empirical quantile estimate is denoted by '0', the members of the sets $A$ and $B$ are denoted by the numbers 1 to 10 (lower index for lower threshold) and 11 to 20 (lower index for lower threshold), respectively.}
	\label{fig03}	
\end{figure}

Figure \ref{fig03} shows the frequency distributions of the competing predictors set $A\cup B$ selected as optimum through $\widehat{Q}_{qs}$, $\widehat{Q}_1$ and $\widehat{Q}_2$ when the data is generated from the mixture model-$(ii): (a)$. It is found that $\widehat{Q}_{qs}$ uniformly selects the predictors with slightly more preference to higher threshold based predictors. Whereas $\widehat{Q}_1$ frequently selects lower thresholds, indicating a preference to the predictors based on a large number of observations to be fitted to the GPD. It is also found that $\widehat{Q}_2$ selects the lower thresholds of both $A$ and $B$ more frequently. It can be said that $\widehat{Q}_2$ has less discriminate power compared to $\widehat{Q}_1$ for distinguishing sets $A$ and $B$. Note that, for the original sample of size $n=7500$, the training sample size for $\widehat{Q}_2$ turns out to more than 5000 giving a reasonably large sample to estimate all the predictors of $A \cup B$. When the training sample size is reduced to 2500, the method tends to select the predictor from the set $A$ (the results are omitted for brevity). 

\begin{table}
	\caption{RMSE of the optimum score predictors $\widehat{Q}_{qs}$, $\widehat{Q}_1$ and $\widehat{Q}_2$ for the quantile at probability level $p^0=(1-\frac{1}{2n})$, based on samples of size $n=7500$ and using $\alpha=\{1, 2, 4, 8\}$. All the results are based on 1000 iterations.}
	\vspace{0.4cm}
	\centering\label{tab-rmse001}
	%\resizebox{\columnwidth}{!}{
	\begin{scriptsize}
		\begin{tabular}{ccccccc@{\hskip 10pt}c@{\hskip 10pt}c@{\hskip 10pt}c}\hline& \multicolumn{3}{c}{AB} \vline& \multicolumn{3}{c}{A}\vline&\multicolumn{3}{c}{B} \\\cline{2-10}\\
			Model   & $\widehat {Q}_{qs}$& $\widehat {Q}_1$&$\widehat {Q}_2$  &  $\widehat {Q}_{qs}$&  $\widehat {Q}_1$&$\widehat {Q}_2$ & $\widehat {Q}_{qs}$& $\widehat {Q}_1$&$\widehat {Q}_2$\\\hline\\
			$(i): (a)$ &0.012   &0.011   &0.012   &0.012  &0.013    &0.013   
			& 0.012   & 0.010   &0.012\\
			$(i): (b)$ &1.171   &0.558   &0.934   &1.172  &1.120    &1.141   
			& 1.170   & 0.473   &0.943 \\
			$(i): (c)$ &585.95 &68.70 &112.95 
			&585.77& 69.21 & 110.87 
			& 586.10 &44.47  &109.15 \\
			$(ii): (a)$&230.92 &87.265  &92.63 &228.92&192.03   &92.81 
			& 228.92 &90.72  &92.62 \\
			$(ii): (b)$&25.175  &411.714 &62.334  &25.15 &28.86  &23.15  
			& 25.15  & 12.67  &22.68\\
			$(iii)$  &231.256 &97.315  &98.983 &229.29& 192.66  &94.58 
			& 229.28 &93.12  &98.36\\
			$(iv)$   &1.114   &1.054   &1.021 
			&1.107  &1.031  &1.027 
			& 1.108   &1.031    &1.027 \\
			[0.1cm]\hline
		\end{tabular}		
	\end{scriptsize}
	%}
\end{table}

We then compare the above three methods with the median of the competing predictors and a randomly chosen predictor from the competing set. As before, we use the predictor set $\{0\}\cup A\cup B$ for estimating the quantile at probability level $p^0=1-\frac{1}{2n}$ and use the tuning parameter $\alpha\in \{1, 2, 4, 8\}$. Table~\ref{tab-rmse002} shows the RMSE of the optimum score predictors $\widehat{Q}_{qs}$, $\widehat{Q}_1$, $\widehat{Q}_2$, along with those of the random and the median prediction. As expected, the performance of the random predictor is worst compared to the other predictors. The predictor $\widehat{Q}_1$ dominates the median prediction except model $(ii): (b)$ and $(iv)$.

\begin{table}
	\caption{RMSE of the optimum score predictors $\widehat{Q}_{qs}$, $\widehat{Q}_1$, $\widehat{Q}_2$, median predictor $\widehat {Q}_{med}$ and random predictors $\widehat {Q}_{rand}$ for the quantile at probability level $p^0=(1-\frac{1}{2n})$, based on samples of size $n=7500$. All the results are based on 1000 iterations.}
	\vspace{0.4cm}
	\centering\label{tab-rmse002}
	%\resizebox{\columnwidth}{!}{
	\begin{scriptsize}
		\begin{tabular}{cccccccc}\\\hline\\
			Model   & $\widehat {Q}_{qs}$& $\widehat {Q}_1$&$\widehat {Q}_2$  &  $\widehat {Q}_{med}$ &$\widehat {Q}_{rand}$\\\hline\\
			
$(i): (a)$ &0.012   &0.011   &0.012   &0.013&0.013  \\
$(i): (b)$ &1.171   &0.558   &0.934   &1.067&1.137 \\
$(i): (c)$ &585.95 &68.70 &112.95 &133.94&169.29\\
$(ii): (a)$&30.92 &87.265  &92.63  &87.17&118.60\\
$(ii): (b)$&25.175  &411.714 &62.334 &16.58&322.22 \\
$(iii)$ &231.256 &97.315  &98.983  &107.93&112.74\\
$(iv)$   &1.114   &1.054   &1.021    &0.989&0.982\\
			[0.1cm]\hline
		\end{tabular}
		
	\end{scriptsize}
	%}
\end{table}

\section{Assessment of quantile predictions at high tails: real data examples}\label{data_analysis}
In this section, we illustrate the proposed scoring methods with two different real data examples: cyber Netflow bytes transfer and daily precipitation. 
\subsection{LANL cyber netflow data}\label{lanl}
The unified Netflow and host event dataset, available from Los Alamos National Laboratory (LANL, https://csr.lanl.gov/data/2017.html), thoroughly described in \cite{Turcotte2018}, is one of the most commonly used datasets in cybersecurity research for anomaly detection with enormous importance in industry and society \citep{Adams2016}. The data comprise records describing communication events between various devices connected to the LANL enterprise. %various aspect of netflow data is throughly described in \cite{Turcotte_2017}.
% It has been shown that flow based techniques have a number of advantages and are successful in detecting a variety of malicious network behaviors \citep{Sperotto2010}. Detection of anomaly using the new edge detection has been investigated by \cite{Metelli_2016}. \cite{Heard_2016} developed a stochastic process based approach for anomaly detection.
The daily Netflow data consisting of hundreds of thousands of records is available for 90 days (day 2 to day 91, starting from a specific epoch time). Each of the flow records (Netflow V9) is an aggregate summary of a bi-directional network communication with the following components: {\it StartTime, EndTime, SrcIP, DstIP, Protocol, SrcPort, DstPort, SrcPackets, DstPackets, SrcBytes} and {\it DstBytes} \citep{Turcotte_2017}. In this work, we analyze one important component of this data set.

\begin{figure}
	\centering
	\includegraphics[height=3in]{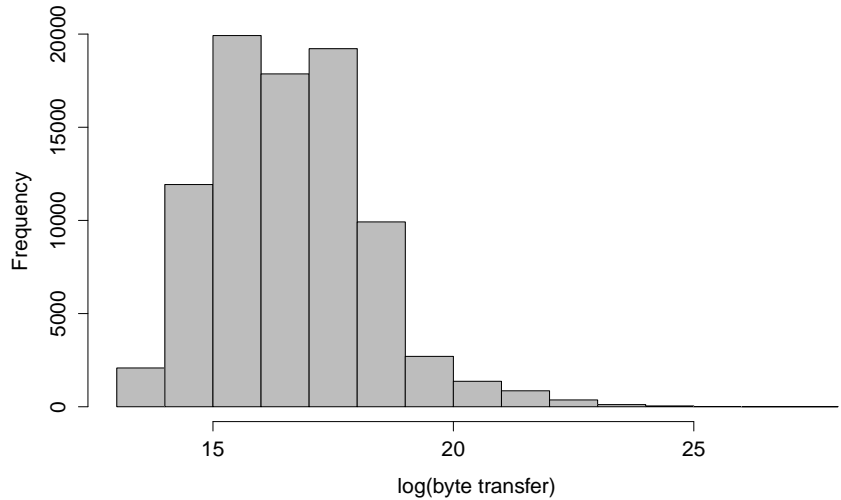}
	\caption[]{Histogram of the total volume of bytes transfer in a second on a particular day.}\label{hist11}	
\end{figure}

As a variable of interest, we consider the total volume of bytes transfer (through an edge) which started in an epoch time. The data is available for every second of the day, so in a day we have $n$=86400 observations. Figure \ref{hist11} shows the histogram of the byte transfer in a typical day (day 3, chosen randomly from the available 90 days), indicating the heavy-tailed nature of the underlying distribution. We consider the target quantile as $p^0=(1-\frac{1}{2n})$ and use the competing set of 21 predictors $\{0\}\cup A \cup B$. As in the simulation study, we consider the same set of values of the tuning parameter $k$ (or equivalently $\alpha$) as $\{3, 5, 7 19\}$.

As the variable of interest, we consider the total volume of bytes transfer. The optimum scoring predictors through methods $\widehat{Q}_{qs}$, $\widehat{Q}_1$ and $\widehat{Q}_2$ turn out to be based on 20th upper order statistics and percentiles with probability levels 0.99467 and 0.996. The optimal predictions of $p^0$th quantile of total bytes transfer using the above three methods are 689.24 gigabytes, 2070.581 gigabytes and 1007.86 gigabytes, respectively.

\subsection{GHCN daily precipitation data}\label{ghcn}
Prediction of extreme high return values of daily precipitation is a common task in hydrological engineering where these values correspond to return periods of 100 or 1000 years based on relatively few years of data \citep{bader2018}. 
Daily precipitation data is available from Global Historical Climatology Network (GHCN) and can be downloaded freely from ftp://ftp.ncdc.noaa.gov/\\pub/data/ghcn/daily/ for tens of thousands of surface weather sites around the world \citep{Menne2012}. To illustrate the proposed methodology, we consider a sample station chosen randomly from the set of stations in one of the US coastal state, California. 

\begin{figure}
	\centering
	\includegraphics[height=3in]{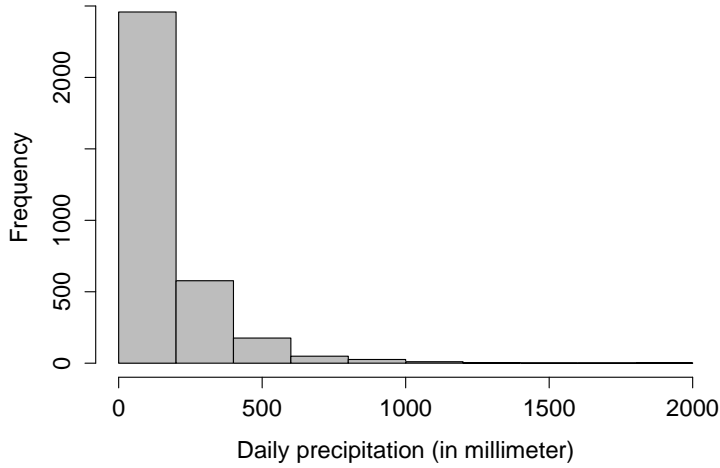}
	\caption[]{Histogram of the daily precipitation during the time period 1945-2010 at a met-station in California.}\label{hist22}    
\end{figure}

We consider precipitation data for the months, November to March for each year for the period 1940 to 2015 (following \cite{bader2018}) with the total number of daily data points as $n=$3303 after excluding zeros and some missing observations. Figure \ref{hist22} shows the histogram of the daily precipitation data, indicating the heavy-tailed nature of the response variable. Here we set our target as to predict the $2n$-observation (150-year) return level, implying $p^0=(1-\frac{1}{2n})$. As a set of predictors, we consider 21 unconditional predictors $\{0\}\cup A\cup B$ considered in Section~\ref{simulation}. As the values of $k$ (or equivalent values of $\alpha$), we take the set of values (3, 5, 7, 17). The optimum predictors using the three methods $\widehat{Q}_{qs}$, $\widehat{Q}_1$ and $\widehat{Q}_2$ are the 20th, fourth and the first member of the competing set and the optimum predictions are 1847.21, 2227.97 and 2727.22 millimetres, respectively. 

\section{Concluding remarks}\label{sec5} 
We have proposed two distribution-free scoring methods for extreme high quantile {\color{black} predictors} to assess their predictive performances. These methods rely on cross-validation by partitioning the sample of $n$ into $k$ folds. The first proposed method evaluates the competing predictors by employing them to predict a series of equally extreme quantiles on a fold of size $\frac{n}{k}$ chosen randomly from the given data set. The predictors are then evaluated by their predictive performance in the other parts of the sample of sizes $(k-1)n/k$. The choice of the number of folds $k$ (obtained through choosing values of $\alpha$) and equally extreme quantile points are obtained by solving two equations for a given tuning parameter set $\boldsymbol{\alpha}$. The second method uses $k-1$ folds, all at a time, to train the predictors and test them on one fold. A numerical study shows the increased efficiency of the proposed methods compared to the conventional quantile scoring method.

{\color{black} The choice of the tuning parameter $\boldsymbol{\alpha}$ is a crucial issue. For a sample of size $n$ and the given target probability, $p_0=1-\frac{1}{2n}$, a set of choices of the tuning parameter $\alpha$ for the two proposed method is given here. We also compare the choice of $\alpha$s in the two proposed methods by fixing the number of folds ($k$). Our simulation study provides a prescription for the choice of the tuning parameter through theoretical justification from quantile regression. This work could be extended by incorporating covariate information in the quantile evaluation and by further investigating the choice of the tuning parameter $\alpha$.}

\begin{acknowledgements} The research of Kaushik Jana is supported by the Alan Turing Institute- Lloyd's Register Foundation Programme on Data-Centric Engineering. The authors also like to thank Dr Tobias Fissler of Vienna University of Economics and Business for many helpful discussions. 
\end{acknowledgements}

\noindent {\bf Statement on Conflict of Interest:}
The authors declare that there is no conflict of interest.

%\begin{acknowledgements}
%\end{acknowledgements}



\bibliographystyle{apalike}
\bibliography{draft}

\end{document}